\documentclass[arxiv,superscriptaddress,twocolumn]{revtex4-1}

\usepackage{lineno,hyperref}

\usepackage{graphicx}
\usepackage{bmpsize}
\usepackage{amsfonts}
\usepackage[usenames, dvipsnames]{color}

\newcommand{\ben}{\begin{eqnarray}}
\newcommand{\een}{\end{eqnarray}}

\newcommand{\bef}{\begin{figure}[h!bt]\centering}
\newcommand{\eef}{\end{figure}}
\newcommand{\bet}{\begin{table}[hbt]\centering}
\newcommand{\eet}{\end{table}}

\begin{document}
\title{Synthesis, magnetization and heat capacity of triangular lattice materials NaErSe$_2$ and KErSe$_2$}
\author{Jie Xing}
\affiliation{Materials Science and Technology Division, Oak Ridge National Laboratory, Oak Ridge, Tennessee 37831, USA}
\author{Liurukara D. Sanjeewa}
\affiliation{Materials Science and Technology Division, Oak Ridge National Laboratory, Oak Ridge, Tennessee 37831, USA}
\author{Jungsoo Kim}
\affiliation{Department of Physics, University of Florida, Gainesville, Florida 32611, USA}
\author{William R. Meier}
\affiliation{Materials Science and Technology Division, Oak Ridge National Laboratory, Oak Ridge, Tennessee 37831, USA}
\author{Andrew F. May}
\affiliation{Materials Science and Technology Division, Oak Ridge National Laboratory, Oak Ridge, Tennessee 37831, USA}
\author{Qiang Zheng}
\affiliation{Materials Science and Technology Division, Oak Ridge National Laboratory, Oak Ridge, Tennessee 37831, USA}
\author{Radu Custelcean}
\affiliation{Chemical Sciences Division, Oak Ridge National Laboratory, Oak Ridge, Tennessee 37831, USA}
\author{G. R. Stewart}
\affiliation{Department of Physics, University of Florida, Gainesville, Florida 32611, USA}
\author{Athena S. Sefat}
\affiliation{Materials Science and Technology Division, Oak Ridge National Laboratory, Oak Ridge, Tennessee 37831, USA}

\date{\today}

\begin{abstract}
In this paper we report the synthesis, magnetization and heat capacity of the frustrated magnets \emph{A}ErSe$_2$(\emph{A}=Na,K) which contain perfect triangular lattices of Er$^{3+}$. The magnetization data suggests no long-range magnetic order exists in \emph{A}ErSe$_2$(\emph{A}=Na,K), which is consistent with the heat capacity measurements. Large anisotropy is observed between the magnetization within the \emph{ab} plane and along the \emph{c} axis of both compounds. When the magnetic field is applied along \emph{ab} plane, anomalies are observed at 1.8 $\mu_B$ in NaErSe$_2$ at 0.2 T and 2.1 $\mu_B$ in KErSe$_2$ at 0.18 T. Unlike NaErSe$_2$, a plateau-like field-induced metamagnetic transition is observed for H$\|$\emph{c} below 1 K in KErSe$_2$. Two broad peaks are observed in the heat capacity below 10 K indicating possible crystal electric field(CEF) effects and magnetic entropy released under different magnetic fields.  All results indicate that \emph{A}ErSe$_2$ are strongly anisotropic, frustrated magnets with field-induced transition at low temperature. The lack of signatures for long-range magnetic order implies that these materials are candidates for hosting a quantum spin liquid ground state.
\end{abstract}

\maketitle

\section{Introduction}
Geometrically frustrated magnets with many novel properties in the low temperature region have been studied for many years\cite{review}. One intriguing phenomenon is the magnetization plateau. An ordered ground state of 120$^\circ$ between each spin was proposed in two-dimensional spin 1/2 triangular lattice Heisenberg antiferromagnet (TLHAF) \cite{TLHAF1}. Due to the degeneration of the energy in the quantum system, the magnetization under the magnetic field is predicted with a plateau at 1/3 of the saturation magnetization in TLHAF \cite{plateau1,plateau2,plateau4}. On the experimental side, the 1/3 plateau is found in triangular lattices materials with varied spin number from 1/2 to 3/2, such as Cs$_2$CuBr$_4$, CuFeO$_2$, RbFe(MoO$_4$)$_2$ and Ba$_3$CoSb$_2$O$_9$\cite{plateaumodel1,plateaumodel2,plateaumodel3,plateaumodel4}. Furthermore, besides the TLHAF, strong spin-lattices coupling could induce a 1/2 magnetization plateau in the geometrical frustrated system\cite{plateaumodel6,plateaumodel7}.

Another important topic relevant to geometrically frustrated magnets is the quantum spin liquid(QSL). The QSL state is a highly entangled quantum state leading to the fractionization of spin degrees of freedom\cite{anderson}. Recently, Majorana fermions were also proposed in the quantum spin liquid system \cite{Kitaev}. Until now several compounds with \emph{S}=1/2 are proposed as the QSL candidates , such as honeycomb iridates \emph{A}$_2$IrO$_3$ (\emph{A}=Na,Li,H$_3$Li,Cu), $\kappa$-(BEDT-TTF)$_2$Cu$_2$(CN)$_3$, EtMe$_3$Sb[Pd(dmit)$_2$]$_2$, and RuCl$_3$\cite{Na2IrO3, A2IrO3, A2IrO32, Na2IrO33, Na2IrO34, Li2IrO3, Na2IrO36, Na2IrO37, Cu2IrO3, Et, RuCl3, RuCl32, RuCl33, RuCl34, RuCl35}.

\begin{table*}
\caption{The crystal structure of the NaErSe$_2$ and KErSe$_2$ phase at 300K. Both samples are belong to the same space group $R\overline{3}m$.}
\begin{tabular}{p{2cm}p{3cm}p{2cm}p{2cm}p{2cm}p{1cm}}
\hline\hline
component  & instrument & \emph{a}({\AA}) & \emph{c}({\AA}) &\emph{z}(Se) & R \\
\hline
NaErSe$_2$   & single crystal XRD   &  4.0784(4) & 20.746(3) &0.25694(1) &0.03 \\
NaErSe$_2$   & powder XRD           &  4.0876(1) &20.7980(6) &0.2562(1) &0.08 \\
KErSe$_2$   & single crystal XRD    &  4.1466(1) &22.743(5) &0.2647(1) &0.02 \\
KErSe$_2$   & powder XRD            &  4.1470(1) &22.7662(8) &0.2649(1) &0.07 \\
\hline
\hline
\end{tabular}
\label{tab.1}
\end{table*}

Recently, both theoretical and experimental results indicate the magnetic rare earth ions located in the geometrically frustrated lattices (e.g., triangular lattice) may also form QSL states \cite{balent,RMP}. Rare earth ions in configurations with an odd number of 4\emph{f} electrons support Kramer doublets which can be treated as an effective spin \emph{J}$_{eff}$=1/2. Compared to transition metals, the spin-orbital coupling in the rare earth system is much stronger and highly anisotropic exchange couplings are expected \cite{soc}. YbMgGaO$_4$ with the YbFe$_2$O$_4$ structure is studied as rare earth QSL candidate. The Yb$^{3+}$ ions construct a triangular layer and are octahedrally coordinated by O$^{2-}$ ions. The heat capacity, thermal conductivity, neutron scattering and muon spin relaxation reveal a possible gapless QSL ground state \cite{Yb1,Yb2,Yb3,Yb4,Yb5,Yb6,Yb7,Yb10}. However, the existence of Ga/Mg disorders may drive the system to another state \cite{Yb1,Yb5,Yb6,gapYb1,gapYb2,gapYb3,gapYb4}. Replacing Yb by Er also reveals possible QSL behavior\cite{Er1,Er2}. This encourages research on rare earth materials with the ideal frustrated triangular structures. Very recently a classic system \emph{A}\emph{RE}\emph{Ch}$_2$(\emph{A}=Alkali metal, \emph{RE}=Rare earth elements, \emph{Ch}= chalcogens) with perfect triangular lattices of rare earth ions were proposed as QSL candidates \cite{QMZhang,NaYbS2,NaYbO2,NaYbO22,NaYbO23}. Comparing to the other frustrated lattice structures, this family with less ions and simple triangular structure suggests less possibilities of impurities or disorders. No structural or magnetic transition was found in the NaYb\emph{Ch}$_2$(\emph{Ch}=O,S,Se) down to 50 mK from heat capacity and magnetization \cite{QMZhang}. Single crystal studies outline the strongly anisotropic, quasi 2D magnetism and \emph{J}$_{eff}$=1/2 state in NaYbS$_2$ \cite{NaYbS2}. Specific heat data consistent with a gapless QSL state was found in NaYbO$_2$ and magnetic field promotes a quantum phase transition above 2 T \cite{NaYbO2,NaYbO22,NaYbO23}. Besides these possible QSL state studies, an anisotropic spin of the rare earth ions in a crystal electric field has been proposed by theoretical calculations\cite{chen1,chen2,chen3}. This system is a large family with diversity originating from rare earth, alkali or transition metal and chalcogen, which could support more novel properties of multiple exchange couplings and crystal field effects. From this aspect, this system is a good platform to study frustrated magnetism of rare earth ions on a triangular lattice.

\begin{figure*}
\includegraphics[width=7in]{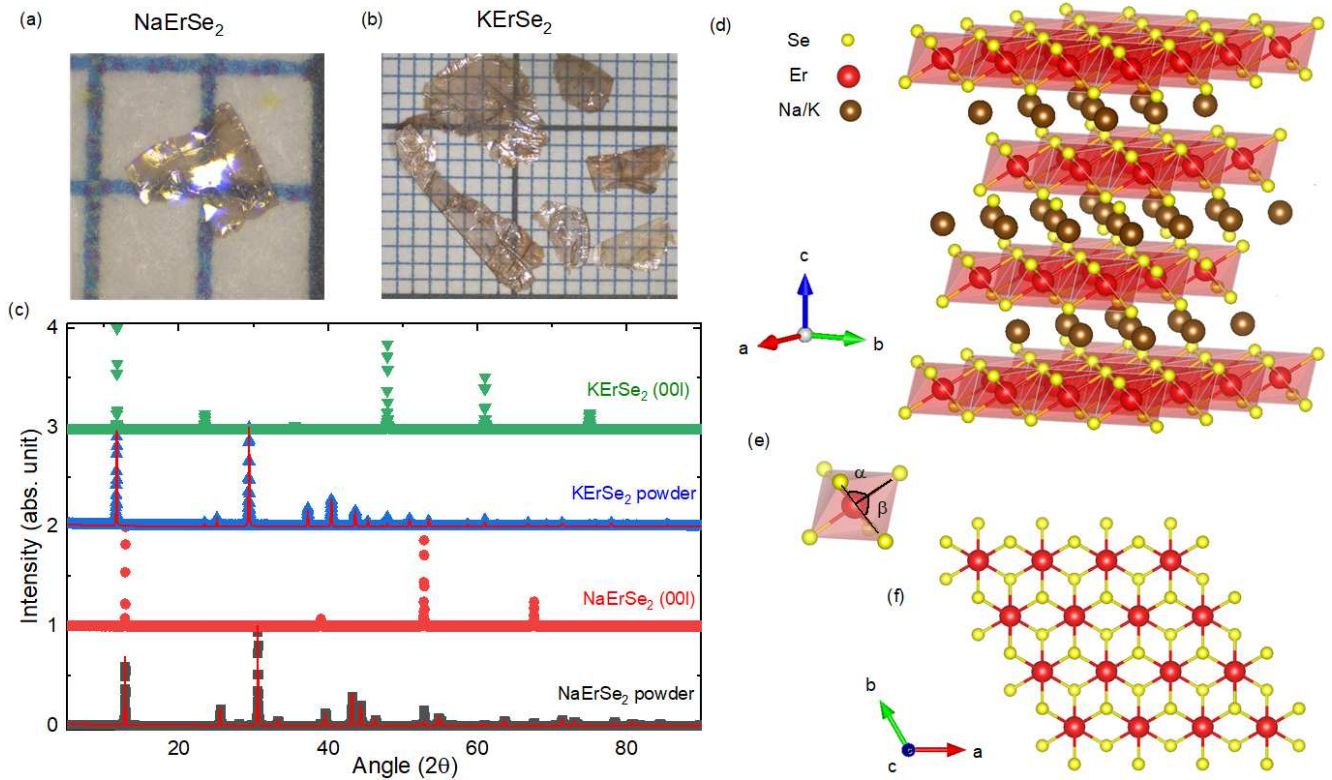}
\caption {(a-b) The crystal picture of \emph{A}ErSe$_2$ (\emph{A}= Na and K) against 1 mm scale. (c) XRD pattern of \emph{A}ErSe$_2$ (\emph{A} = Na and K) powder and crystals on (00l) direction. The red lines show the refinement result of the two compounds. (d) The schematic structure of \emph{R}\={3}\emph{m} \emph{A}ErSe$_2$ (A = Na and K).  The red edge-sharing distorted ErSe$_6$ octahedra construct the triangular layer between the Alkali ions. (e) The Se-Er-Se angle of distorted ErSe$_6$ octahedra. (f) The ideal triangular Er$^{3+}$ layer in the \emph{ab} plane. The nearest neighbour Er$^{3+}$ ions are connected by two Se ions.}
\label{fig1}
\end{figure*}

Until now all the measurements in this \emph{A}\emph{RE}$Ch_2$ system focused on the \emph{A}=Na and \emph{RE}=Yb$^{3+}$ \cite{QMZhang,NaYbS2,NaYbO2,NaYbO22,NaYbO23}. The fruitful results prompted us to undertake an investigation of the physical properties in \emph{A}ErSe$_2$ (\emph{A}=Na, K) systems. In this paper, we report the synthesis of Na/KErSe$_2$ single crystals and survey the magnetization and heat capacity in this system. Besides the prior reported NaErSe$_2$ and NaLuSe$_2$, we synthesize both powder and single crystal of new KEr(Lu)Se$_2$. Good crystallinity is confirmed by the powder X-ray diffraction (XRD) and single crystallines XRD. No site vacancies or impurities are found in the crystal. No magnetic transition is found down to 0.42 K in the magnetization and heat capacity characterization for small applied magnetic fields. Large anisotropy is exposed between H$\|$\emph{ab} and H$\|$\textit{c}. A tiny slope change in the isothermal magnetization is found only for H$\|$\emph{ab} at low temperature. The moment where this occurs is 1.8 $\mu_B$ and 2.1 $\mu_B$ in NaErSe$_2$ and KErSe$_2$. KErSe$_2$ shows a clear plateau-like metamagnetic transition with m=2.3 $\mu_B$ at H$\|$\emph{c} below 1 K. The heat capacity shows two broad features without any $\lambda$ anomaly above 0.4 K at 0 T. All these physical properties indicate the absence of long range order and existence of short-range order in these materials.

\section{Materials and Methods}
Transparent millimeter-scale \emph{A}ErSe$_2$ and \emph{A}LuSe$_2$ single crystals were synthesized by the two-step method. Pictures of the crystals are shown in Fig.1(a-b). At first Na chunks (Alfa Aesar 99.9$\%$), K chunks (Alfa Aesar 99.9$\%$), Er chunks (Alfa Aesar 99.99$\%$) and Se chunks (Alfa Aesar 99.99$\%$) were mixed by the stoichiometric ratio \emph{A}:Er(Lu):Se=1.1:1:2. The alkali element excess 10$\%$ due to the slight reaction with the inner wall of the silica tube. The whole mixture was loaded into a carbon crucible and sealed in the silica tube under vacuum. The ampules were slowly heated to 220$^{\circ}$C and held for 24 h. Then we heated them up to 900$^{\circ}$C and held for 3 days. After the first reaction, the samples cooled down to room temperature over 24 h. The reacted compounds were ground and rinsed with deionized water and acetone. This process gets rid of the multiple binary phases and excess alkali metal. The pure \emph{A}ErSe$_2$ powders are confirmed by powder XRD, as shown in Fig. 1(c). The second step is the flux synthesis. We used NaCl and KCl as the flux for the NaEr(Lu)Se$_2$ and KEr(Lu)Se$_2$, respectively. The mass ratio between the flux and the powder is 10:1. The mixture was sealed in the silica tube under vacuum. The ampoules were heated up to 850$^\circ$C and held for 2 weeks. The compounds were taken out after the fast furnace cooling and leached out in the deionized water and acetone. The sizable, plate-shape single crystals were found and the thickness is related to the dwell time at the second step. No impurities peaks and sharp (00l) manifest good quality for both compounds, as shown in the Fig. 1(c). The slow cooling rate do not exhibit obvious relation with the size of the crystals.

Magnetic properties were measured in Quantum Design (QD) Magnetic Properties Measurement System (MPMS3) with iHe3 option. Due to the shape of these crystals, we stacked several crystals along \emph{c} with mass ~3 mg and used them in the regular magnetic measurement above 2 K. Single crystals of \emph{A}ErSe$_2$(\emph{A}= Na and K) around 0.5-1 mg with as-grown (00l) surfaces were used in the $^3$He option of the MPMS3. Demagnetization factors are considered as N=1 in the isothermal magnetization at H$\|$c. Temperature dependent heat capacity was measured in the QD Physical Properties Measurement System (PPMS) with $^3$He option using the relaxation technique at different applied fields along the c-axis. Single crystal X-ray diffraction(XRD) measurement was performed by Bruker Apex single-crystal X-ray diffractometer. Powder XRD was done by a PANalytical X'pert Pro diffractometer equipped with an incident beam monochrometer (Cu K$\alpha_1$
radiation) at room temperature.

\section{Results and Discussion}

Fig.1(d) shows the schematic structure of \emph{A}ErSe$_2$ (\emph{A}=Na, K with space group \emph{R}\={3}\emph{m}). The perfect triangular layers of magnetic Er$^{3+}$ ions are isolated by the nonmagnetic Na/K layers suggesting anisotropic and frustrated magnetism in this system. The crystal structure parameters of  \emph{A}ErSe$_2$ (\emph{A}=Na, K) obtained from single crystal XRD and powder XRD are shown in Table 1. Both methods yield similar results from the refinements. In this paper we use the powder parameters for discussion, avoiding the possible effect of the stress in the single crystals. Due to the large radius of K$^+$, the distances between each layer are 6.932 {\AA} in NaErSe$_2$ and 7.588 {\AA} in KErSe$_2$. The distance between the nearest neighbor Er$^{3+}$ ions also extended from 4.087 {\AA} to 4.147 {\AA} by replacing K$^+$ from Na$^+$. Although the frustrated interactions exist in the triangular lattice, the interlayer interaction may induce the magnetic transition at the low-temperature region\cite{plateaumodel4}. Therefore, KErSe$_2$ could show more two-dimensional charactor corresponding to the larger \emph{c}/\emph{a} ratio in the structure. In addition, due to the 8\% larger \emph{c}/\emph{a} ratio in KErSe$_2$, the distortion of ErSe$_6$ octahedra is stronger than the NaErSe$_2$. The Se-Er-Se angle $\alpha$ and $\beta$ in Fig.1(e) change from 91.51$^\circ$ and 88.49$^\circ$ in NaErSe$_2$ to 93.13$^\circ$ and 86.87$^\circ$ in KErSe$_2$. The change in local crystalline environment inspires us to investigate how the properties of these two compounds compare.

\begin{figure}
\includegraphics[width=3.5in]{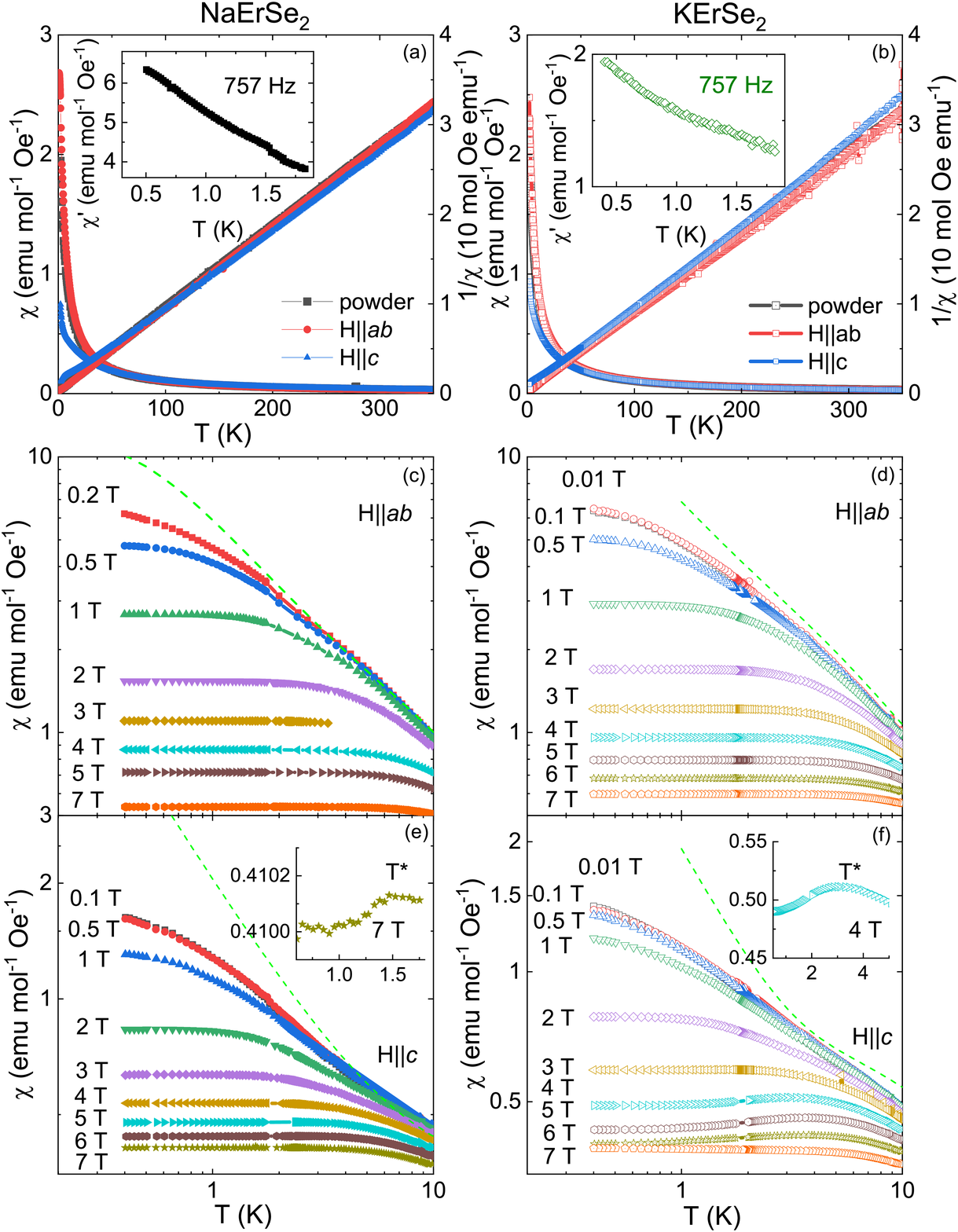}
\caption {(a-b) Temperature dependence of magnetic susceptibility and inversed magnetic susceptibility from 2 K to 350 K for AErSe$_2$ powder and crystal along \emph{ab} plane and \emph{c} axis. The inset show the ac susceptibility from 0.4K to 1.8K for AErSe$_2$. (b)-(f) Temperature dependence of magnetic susptibility in NaErSe$_2$ and KErSe$_2$ crystal with H$\|$\emph{ab} or H$\|$\emph{c}. The inset shows the temperature dependence of magnetic susceptibility at high magnet field. T* is the temperature of the maximum in each magnetic field. The light green dash lines are the fitting from CEF at 0.2 T in (c) and 0.1 T in (d-f).}
\label{Fig2}
\end{figure}

Fig.~2(a-b) presents the temperature dependence of magnetic susceptibility $\chi=M/H$ and inverse magnetic susceptibility from 2 K to 350 K for the powder and single crystals. The anisotropic behaviour appears below 40 K. The magnetization of the single crystals along \emph{H}$\|$\emph{ab} and \emph{H}$\|$\emph{c} is well matched with the result of the powder above 40 K.  No visible features are found above 2 K suggesting there are no magnetic or structural transitions. The inset shows the AC susceptibility of NaErSe$_2$ and KErSe$_2$ from 0.4 K to 1.8 K. No hint of spin freezing or long-range magnetic order appears above 0.4 K. These results all suggest that the perfect triangular lattice with Er$^{3+}$ is stable at low temperature which is consistent with other Yb compounds in this system\cite{QMZhang, NaYbS2, NaYbO2, NaYbO22, NaYbO23}. No structure transition or distortion in Er$^{3+}$ imply that this system is a steady platform to investigate novel magnetic properties in the frustrated system. The Curie-Weiss fitting of the powder above 200 K yield a Curie-Weiss temperature $\theta_{CW}$=-10.9 K and effective moment $\mu_{eff}$=9.5 $\mu_B$ in NaErSe$_2$ and $\theta_{CW}$=-8 K, $\mu_{eff}$=9.5 $\mu_B$ in KErSe$_2$, respectively. The effective moments of these two compounds match the theoretical value of 9.6 $\mu_B$/Er$^{3+}$ for free ions. The negative value of $\theta_{CW}$ indicates the dominant antiferromagnetic interaction between the Er$^{3+}$ ions in the triangular lattice. Curie-Weiss fitting is also applied in both samples from 10 K to 30 K: $\theta_{CW}$=-4.3 K, $\mu_{eff}$= 9.4 $\mu_B$ for NaErSe$_2$ and $\theta_{CW}$=-3.8 K, $\mu_{eff}$= 9.4 $\mu_B$ for NaErSe$_2$. The change of $\theta _{CW}$ may be caused by the thermal population of CEF levels.

Fig. 2(c)-(f) present the temperature dependence of magnetic susceptibility under different magnetic fields for NaErSe$_2$ and KErSe$_2$. When the temperature decreases below 10 K, \emph{M(T)} deviates from the Curie-Weiss behavior. The relation observed in this region is \emph{M} $\propto$ \emph{T}$^{-0.4}$. It is consistent with that recently observed in Os$_{0.55}$Cl$_2$, which displays many physical properties similar to those observed in quantum spin liquid candidates\cite{micheal}. It worth noting that there is no significant long-range magnetic transition above 0.4 K from low magnetic field DC susceptibility and AC susceptibility in zero applied DC field.

The temperature-independent regions under different magnetic fields are found in these two compounds with \emph{H}$\|$\emph{ab}. However, when the magnetic field is along the \emph{c} axis, the magnetic susceptibility reaches a maximum and then decreases with temperature. We choose the starting point of d$\chi$/d$T$(T)=0 as \emph{T}* for \emph{H}$\|$\emph{c}. The subtle hump feature for \emph{H}$\|$\emph{c} in NaErSe$_2$ at 7 T is shown in the inset of Fig.~2(e). The magnitude of the $\Delta \chi$ is only around 0.02\% and is challenging to distinguish at the low magnetic field. The similar but remarkable situation also appears in KErSe$_2$. The magnitude of $\Delta \chi$ is 7\% which is much larger than NaErSe$_2$. This maximum in $\chi$ is observed as high as 3 K and is present for fields between approximately 4 and 6 T. The origin of this feature may be the population of the CEF states, long-range order or short-range magnetic interaction between Er$^{3+}$ ions. We could rule out the long-range magnetic order since no $\lambda$ anomaly is found in the same temperature range in specific heat measurements (Fig.~4). The mixture of the CEF states with magnetic fields in KErSe$_2$ could potentially contribute to the broad peak in the temperature dependence of magnetization.

\begin{figure}
\includegraphics[width=3.6in]{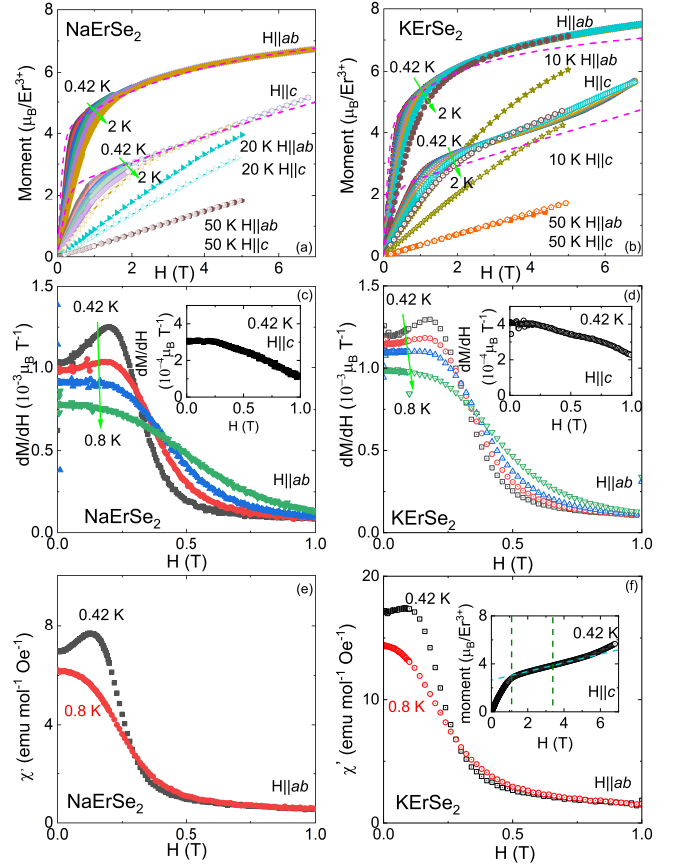}
\caption {(a) Isothermal magnetization of NaErSe$_2$ up to 7 T from 0.42 K to 50 K at \emph{H}$\|$\emph{ab} and \emph{H}$\|$\emph{c}. The pink dash lines show the fitting from the CEF. (b) Isothermal magnetization of KErSe$_2$ up to 7 T from 0.42 K to 50 K at \emph{H}$\|$\emph{ab} and \emph{H}$\|$\emph{c}. The pink dash lines show the fitting from the CEF at 0.42 K. (c-d) Derivation of isothermal magnetization of NaErSe$_2$ and KErSe$_2$ from 0 T to 1 T at \emph{H}$\|$\emph{ab}. Inset: Derivation of isothermal magnetization of NaErSe$_2$ and KErSe$_2$ from 0 T to 1 T along \emph{H}$\|$\emph{c} at 0.42 K. (e-f) AC susceptibility of NaErSe$_2$ and KErSe$_2$. Inset: The isothermal magnetization of KErSe$_2$. The dash line indicates the moment and field range of the plateau.} \label{Fig3}
\end{figure}

The isothermal magnetization is sensitive to the transition induced by the magnetic field. We measured the isothermal magnetization from 0.42 K to 50 K for NaErSe$_2$ and KErSe$_2$ up to 7 T, as shown in Fig. 3 (a) and Fig. 3 (b). The magnetic moments are not saturated at 0.42 K and 7 T for either compound and orientation. Large anisotropy is found in both of them below 50 K. If we linear extrapolated high field data to 0 T, we could get strong anisotropy moments for NaErSe$_2$: \emph{m}$_{ab}$=5.6 $\mu _B$ and \emph{m}$_{c}$=2.1 $\mu _B$. The anisotropic values are lower than the expected value of Er$^{3+}$ which may be affected by the CEF. A subtle change of slope is found at the low magnetic field around 0.2 T presenting obviously by dM/dH in Fig.~3(c). This subtle feature appears only below 0.6 K with the moment 1.8 $\mu_B$, which is close to 1/3 of \emph{m}$_{ab}$=5.6 $\mu _B$ in NaErSe$_2$. The similar subtle hump is also found in the KErSe$_2$ when H $\|$ ab, as shown in Fig.~3(d) as well. The moment around 2.1 $\mu_B$ is even close to 1/3 of the extended moment \emph{m}$_{ab}$=6.2 $\mu _B$ in KErSe$_2$. Because there is no sublet kink near the same magnetic fields at H$\|$\emph{c} in the inset of Fig.~3(c-d), we can exclude the influence of the impurities or other isotropic impacts. The possible phase transition or population of CEF states could induce the slope change. To confirmed this feature, we also measured AC susceptibility at 757 Hz below 1 K along \emph{H} $\|$\emph{ab} and \emph{H}$\|$\emph{c}. The similar broad peaks are located only at \emph{H}$\|$\emph{ab} as shown in Fig.~3(e-f). The magnetic fields of the maximum are almost identical with the DC magnetization in Fig.~3(c). Besides these features at \emph{H}$\|$\emph{ab}, an apparent plateau with upturn at high field is recognized only in the KErSe$_2$ at \emph{H}$\|$\emph{c}, as shown in Fig. 3(b). The field range of the plateau is 2 T to 4.5 T indicating by the vertical green lines in the inset of Fig.~3(f). The linear extended dash line indicates 2.3 $\mu _B$. The length of plateau decreases as increasing the temperature and finally fade out to curved behavior at 2K. It is impressive to find different features along two directions implying the large anisotropic CEF state of Er$^{3+}$ and possible short-range interaction between Er$^{3+}$. In addition, it is worth noting that the compound is not long magnetic ordering in this temperature range even with the field induced plateau-like transition. This behavior is rare in the frustrated system.

\begin{figure}
\includegraphics[width=3.6in]{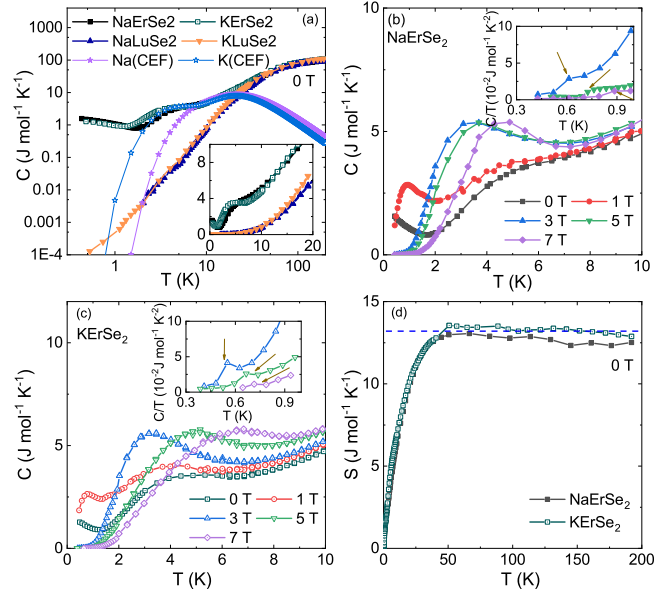}
\caption {(a) The temperature dependence of heat capacity of  NaErSe$_2$, KErSe$_2$,  NaLuSe$_2$ , KLuSe$_2$ and CEF fitting from 0.4 K to 150 K. The inset present the heat capacity of  NaErSe$_2$ and KErSe$_2$ below 20 K. (b) Temperature dependence of C of NaErSe$_2$ under different magnetic fields. The inset shows the tiny peaks at 3 T, 5 T, 7 T. (c) Temperature dependence of C of KErSe$_2$ under different magnetic fields. The inset show the tiny peaks at 3 T, 5 T, 7 T. (d) Temperature dependence of the magnetic entropy of NaErSe$_2$ and KErSe$_2$.
}\label{fig4}
\end{figure}

To investigate the magnetic entropy at low temperature, we measure the heat capacity of NaErSe$_2$ and KErSe$_2$ down to the 0.4 K, as shown in Fig.~4(a). There is no discernible $\lambda$ anomaly down to 0.4 K which agrees with the magnetization measurement. We synthesize nonmagnetic compounds NaLuSe$_2$ and KLuSe$_2$ to estimate the phonon contribution, which are also presented in Fig. 4(a). The magnetic entropy release from high temperature around 50 K in both compounds. This temperature is much higher than the Yb-112 compounds indicating the strong CEF with Er$^{3+}$. Two broad peaks are observed in both samples from 0.4 K to 10 K. The multi-level Schottky or the short-range magnetic order may cause these peaks in the heat capacity. The low-temperature peak moves to high temperature with the magnetic field. Due to the low-temperature peak overlaps with the high-temperature peak above 3 T, it is hard to distinguish these two peaks. The combined high-temperature peak does not change the magnitude, which strongly suggests the Schottky effects from the CEF above 3 T. However, the magnitude of the peak changes below 3 T and the high temperature peak shift slightly to low temperature at 1 T. These features indicate the entropy changes across 3 T by the possible mixtured CEF states or short range effects. The larger magnetic field($>$3 T) could induce a tiny anomaly below 0.8 K, as shown in the inset of Fig.~4(b-c). The temperature of this tiny transitions increases with enhancing the magnetic field which violates antiferromagnetic transition. This feature indicates the possible subtle interaction of CEF states with different fields. At last, we want to estimate the entropy release in NaErSe$_2$ and KErSe$_2$ at 0 T. The entropy from 0.4 K to 200 K is close to 13 J mol$^{-1}$ K$^{-1}$ in both compounds. This value is larger than the Rln2 and much smaller than the expected 23 J mol$^{-1}$ K$^{-1}$ for Er$^{3+}$. This indicates large entropy releases below 0.4 K or possible high energy CEF exits.

We use a CEF model to fit the magnetization vs temperature and field data to examine the role of single ion effects in these two systems. CEF Hamiltonian of \emph{D}$_{3d}$ symmetry is $H^{CEF}_{D_{3d}}=B^0_2\hat{O}^0_2+B^0_4\hat{O}^0_4+B^3_4\hat{O}^3_4+B^0_6\hat{O}^0_6+B^3_6\hat{O}^3_6+B^6_6\hat{O}^6_6$. This point symmetry splits the \emph{J}=15/2 states into the 3 $\Gamma_{45}^+$ doublets and 5 $\Gamma_{6}^+$ doublets. The fitted curves show good agreements with the magnitude of the M(H) data and the high temperature data. The ground state doublet is reliably $\Gamma_{6}^+$ with the $\Gamma_{45}^+$ as the first excited state around 10-25 K for both samples. The poor agreements at low fields or low temperature revealing that the CEF model is not sufficient alone. The fitting shows a plateau feature could be induced by the mixture of CEF states with magnetic field. Considering the two broad peaks overlap and form Schottky-like peak between 1 T and 3 T in the heat capacity, the plateau for H$\parallel$c in KErSe$_2$ is suggested as the metamagnetic transition from CEF. The short-range interaction between Er$^{3+}$ ions or low-lying CEF levels could cause the anomalies at low field for H$\parallel$ab and the deviation below 2 K. A detailed study of CEF is expected to be done in neutron scattering measurement.

\section{Conclusion}
In conclusion, we present the experimental results of magnetization and heat capacity measurement on \emph{A}ErSe$_2$ (\emph{A}=Na and K) single crystals. Large anisotropy is found below 40 K along two directions \emph{H}$\|$\emph{ab} and \emph{H}$\|$\emph{c}. The isothermal magnetization shows a subtle kink feature only \emph{H}$\|$\emph{ab} at low temperature in both compounds. KErSe$_2$ also shows a wide plateau-like metamagnetic transition for \emph{H}$\|$\emph{c}. The heat capacity show two broad peaks below 10 K at zero field. All results indicate that \emph{A}ErSe$_2$ are strongly anisotropic, frustrated magnets with field-induced transitions at low temperatures. No long-range magnetic order implies that these materials are candidates for hosting a quantum spin liquid ground state.

\section{acknowledgements}

The research is supported by the U.S. Department of Energy (DOE), Office of Science, Basic Energy Sciences (BES), Materials Science and Engineering Division. The X-ray diffraction analysis by RC was supported by the US Department of Energy, Office of Science, Basic Energy Sciences, Chemical Sciences, Geosciences, and Biosciences Division. Work at Florida by J. S. Kim and G. R. S. supported by the US Department of Energy, Basic Energy Sciences, contract no. DE-FG02-86ER45268

Notice: This manuscript has been authored by UT-Battelle, LLC under Contract No. DE-AC05-00OR22725 with the U.S. Department of Energy. The United States Government retains and the publisher, by accepting the article for publication, acknowledges that the United States Government retains a non-exclusive, paid-up, irrevocable, world-wide license to publish or reproduce the published form of this manuscript, or allow others to do so, for United States Government purposes. The Department of Energy will provide public access to these results of federally sponsored research in accordance with the DOE Public Access Plan (http://energy.gov/downloads/doe-public-access-plan).


\begin{thebibliography}{99}
\bibitem{review} J.F. Sadoc, R. Mosseri, Geometrical Frustration, Cambridge University Press, Cambridge, 1999.
\bibitem{TLHAF1} D. A. Huse and V. Elser, Phys. Rev. Lett. \textbf{60}, 2531 (1988).
\bibitem{plateau1} H. Nishimori and S. Miyashita, J. Phys. Soc. Jpn. \textbf{55}, 4448 (1986).
\bibitem{plateau2} A.V. Chubukov and D. I. Golosov, J. Phys. Condens. Matter \textbf{3}, 69 (1991).
\bibitem{plateau4} J. Alicea, A.V. Chubukov, and O. A. Starykh, Phys. Rev. Lett. \textbf{102}, 137201 (2009).
\bibitem{plateaumodel1} T. Ono, H. Tanaka, H. Aruga Katori, F. Ishikawa, H. Mitamura, and T. Goto, Phys. Rev. B \textbf{67}, 104431 (2003).
\bibitem{plateaumodel2} N. Terada, Y. Narumi, K. Katsumata, T. Yamamoto, U. Staub, K. Kindo, M. Hagiwara, Y. Tanaka, A. Kikkawa, H. Toyokawa, T. Fukui, R. Kanmuri, T. Ishikawa, and H. Kitamura, Phys. Rev. B \textbf{74}, 180404 (2006).
\bibitem{plateaumodel3} L. E. Svistov, A. I. Smirnov, L. A. Prozorova, O. A. Petrenko, L. N. Demianets, and A. Y. Shapiro, Phys. Rev. B \textbf{67}, 094434 (2003).
\bibitem{plateaumodel4} Y. Shirata, H. Tanaka, A. Matsuo, and K. Kindo, Phys. Rev. Lett. \textbf{108}, 057205 (2012).
\bibitem{plateaumodel6} A. Miyata, H. Ueda, Y. Ueda, H. Sawabe, and S. Takeyama, Phys. Rev. Lett. \textbf{107}, 207203 (2011).
\bibitem{plateaumodel7} Y. Okamoto, D. Nakamura, A. Miyake, S. Takeyama, M. Tokunaga, A. Matsuo, K. Kindo, and Z. Hiroi, Phys. Rev. B \textbf{95}, 134438 (2017).
\bibitem{anderson} P.W. Anderson, Mater. Res. Bull. \textbf{8}, 153 (1973).
\bibitem{Kitaev} A. Kitaev, Ann. Phys. (NY) \textbf{321}, 2 (2006)
\bibitem{Na2IrO3} Y. Singh and P. Gegenwart, Phys. Rev. B \textbf{82}, 064412 (2010)
\bibitem{A2IrO3} Y. Singh, S. Manni, J. Reuther, T. Berlijn, R. Thomale, W. Ku, S. Trebst, and P. Gegenwart, Phys. Rev. Lett. \textbf{108}, 127203 (2012)
\bibitem{A2IrO32} J. Chaloupka, G. Jackeli, and G. Khaliullin, Phys. Rev. Lett. \textbf{105}, 027204 (2010)
\bibitem{Na2IrO33} I. I. Mazin, H. Jeschke, K. Foyevtsova, R. Valenti, and D. Khomskii, Phys. Rev. Lett. \textbf{109}, 197201(2012).
\bibitem{Na2IrO34} F. Ye, S. Chi, H. B. Cao, B. C. Chakoumakos, J. A. FernandezBaca, R. Custelcean, T. F. Qi, O. B. Korneta, and G. Cao, Phys. Rev. B \textbf{85}, 180403(R) (2012)
\bibitem{Li2IrO3}  T. Takayama, A. Kato, R. Dinnebier, J. Nuss, H. Kono, L. Veiga, G. Fabbris, D. Haskel, and H. Takagi, Phys. Rev. Lett. \textbf{114}, 077202 (2015).
\bibitem{Na2IrO36} S. Chun, J. Kim, J. Kim, H. Zheng, C. Stoumpos, C. Malliakas, J. Mitchell, K. Mehlawat, Y. Singh, Y. Choi, T. Gog, A. Al-Zein, M. Sala, J. Krisch, M. Chaloupka, G. Jackeli, G. Khaliullin, and B. J. Kim, Nature Phys. \textbf{11}, 462 (2015)
\bibitem{Na2IrO37}  K. Kitagawa, T. Takayama, Y. Matsumoto, A. Kato, R. Takano, Y. Kishimoto, S. Bette, R. Dinnebier, G. Jackeli and H. Takagi, Nature \textbf{554}, 341 (2018).
\bibitem{Cu2IrO3}  M. Abramchuk, C. Ozsoy-Keskinbora, J. W. Krizan, K. R. Metz, D. C. Bel, F. Tafti, J. Am. Chem. Soc. \textbf{139}, 15371 (2017)
\bibitem{Et} T. Itou, A. Oyamada, S. Maegawa, M. Tamura, R. Kato, Phys. Rev. B \textbf{77}, 104413 (2008).
\bibitem{RuCl3} A. Banerjee, C. Bridges, J. Yan, A. Aczel, L. Li, M. Stone, G. Granroth, M. Lumsden, Y. Yiu, J. Knolle, S. Bhattacharjee, D. Kovrizhin, R. Moessner, D. Tennant, D. Mandrus, and S. Nagler, Nat. Mater., \textbf{15}, 733 (2016).
\bibitem{RuCl32} A. Banerjee, J. Yan, J. Knolle, C. Bridges, M. Stone, M. Lumsden, D. Mandrus, D. Tennant, R. Moessner, and S. Nagler, Science \textbf{356}, 1055 (2017).
\bibitem{RuCl33} H. Cao, A. Banerjee, J. Yan, C. Bridges, M. Lumsden, D. Mandrus, D. Tennant, B. Chakoumakos, and S. Nagler, Phys. Rev. B \textbf{93}, 134423 (2016).
\bibitem{RuCl34}  K. Ran, J. Wang, W. Wang, Z. Dong, X. Ren, S. Bao, S. Li, Z. Ma, Y. Gan, Y. Zhang, J. Park, G. Deng, S. Danilkin, S. Yu, J. Li, and J. Wen, Phys. Rev. Lett. \textbf{118}, 107203 (2017).
\bibitem{RuCl35}  I. Leahy, C. Pocs, P. Siegfried, D. Graf, S. Do, K. Choi, B. Normand, and M. Lee, Phys. Rev. Lett. \textbf{118}, 187203 (2017).
\bibitem{balent} L. Balents, Spin liquids in frustrated magnets, Nature \textbf{464}, 199 (2010).
\bibitem{RMP} J. S. Gardner, M. J. P. Gingras, and J. E. Greedan, Rev. Mod. Phys. \textbf{82}, 53 (2010).
\bibitem{soc} F. Y Li, Y. D. Li, Y. Yu, A. Paramekanti and G. Chen, Phys. Rev. B, \textbf{95}, 085132 (2017).
\bibitem{Yb1} Y. Li, H. Liao, Z. Zhang, S. Li, F. Jin, L. Ling, L. Zhang, Y. Zou, L. Pi, Z. Yang, J. Wang, Z. Wu, and Q. Zhang, Sci. Rep. \textbf{5}, 16419 (2015).
\bibitem{Yb2} Y. Li, G. Chen, W. Tong, L. Pi, J. Liu, Z. Yang, X. Wang, and Q. Zhang, Phys. Rev. Lett. \textbf{115}, 167203 (2015).
\bibitem{Yb3} Yuesheng Li, Devashibhai Adroja, Pabitra K. Biswas, Peter J. Baker, Qian Zhang, Juanjuan Liu, Alexander A. Tsirlin, Philipp Gegenwart, and Qingming Zhang, Phys. Rev. Lett. 117, 097201 (2016).
\bibitem{Yb4} Y. Shen, Y. Li, H. Wo, Y. Li, S. Shen, B. Pan, Q. Wang, H. C. Walker, P. Steffens, M. Boehm, Y. Hao, D. L. Quintero-Castro, L. W. Harriger, M. D. Frontzek, L. Hao, S. Meng, Q. Zhang, G. Chen, and J. Zhao, Nature \textbf{540}, 559562 (2016).
\bibitem{Yb5} J. Paddison, M. Daum, Z. Dun, G. Ehlers, Y. Liu, M. B. Stone, H. Zhou, and M. Mourigal, Nature Physics \textbf{13}, 117122 (2017).
\bibitem{Yb6} X. Zhang, F. Mahmood, M. Daum, Z. Dun, J. Paddison, N. Laurita, T. Hong, H. Zhou, N. Armitage, and M. Mourigal, Phys. Rev. X \textbf{8}, 031001 (2018).
\bibitem{Yb7} Y. Li, D. Adroja, R. Bewley, D. Voneshen, A. Tsirlin, P. Gegenwart, and Q. Zhang, Phys. Rev. Lett. \textbf{118}, 107202 (2017).
\bibitem{Yb10} W. Steinhardt, Z. Shi, A. Samarakoon, S. Dissanayake, D. Graf, Y. Liu, W. Zhu, C. Marjerrison, C. Batista, S. Haravifard, arXiv:1902.07825 (2019).
\bibitem{gapYb1} Y. Shen, Y. Li, H. C. Walker, P. Steffens, M. Boehm, X. Zhang, S. Shen, H. Wo, G. Chen, and J. Zhao, Nat. Comm. \textbf{9}, 4138 (2018).
\bibitem{gapYb2} Z. Zhu, P. A. Maksimov, Steven R. White, and A. L. Chernyshev, Phys. Rev. Lett. \textbf{119}, 157201 (2017).
\bibitem{gapYb3} Z. Zhu, P. A. Maksimov, Steven R. White, and A. L. Chernyshev, Phys. Rev. Lett. \textbf{120}, 207203 (2018).
\bibitem{gapYb4} I. Kimchi, A. Nahum, and T. Senthil, Phys. Rev. X \textbf{8}, 031028 (2018).
\bibitem{Er1} F. Cevallos, K. Stolze, R. Cava, Sol. Stat. Comm. 278, 5 (2018).
\bibitem{Er2} Y. Cai, C. Lygouras, G. Thomas, M. N. Wilson, J. Beare, D. R. Yahne, K. Ross, Z. Gong, Y. J. Uemura, H. A. Dabkowska, G. M. Luke, arXiv 1905.12798 (2019).
\bibitem{chen1} Y. Li, X. Wang, and G. Chen, Phys. Rev. B \textbf{94}, 035107 (2016).
\bibitem{chen2} Y. Li, X. Wang, and G. Chen, Phys. Rev. B \textbf{94}, 201114 (2016).
\bibitem{chen3} C. Liu, Y. Li, and G. Chen, Phys. Rev. B \textbf{98}, 045119 (2018).
\bibitem{QMZhang} W. Liu, Z. Zhang , J. Ji, Y. Liu, J. Li, X. Wang, H. Lei, G. Chen, and Q. Zhang, Chinese Physics Letters \textbf{35}, 117501 (2018).
\bibitem{NaYbS2} M. Baenitz, Ph. Schlender, J. Sichelschmidt, Y. A. Onykiienko, Z. Zangeneh, K. M. Ranjith, R. Sarkar, L. Hozoi, H. C. Walker, J.-C. Orain, H. Yasuoka, J. van den Brink, H. H. Klauss, D. S. Inosov, and Th. Doert, Phys. Rev. B 98, 220409 (2018).
\bibitem{NaYbO2} L. Ding, P. Manuel, S. Bachus, F. Gru$\beta$ler, P. Gegenwart, J. Singleton, R. D. Johnson, H. C. Walker, D. T. Adroja, A. D. Hillier, A. A. Tsirlin, arXiv:1901.07810 (2019).
\bibitem{NaYbO22} K. M. Ranjith, D. Dmytriieva, S. Khim, J. Sichelschmidt, D. Ehlers, H. Yasuoka, J. Wosnitza, A. A. Tsirlin, H. Kuhne, M. Baenitz, Phys. Rev. B 99, 180401 (2019).
\bibitem{NaYbO23} M. Bordelon, E. Kenney, T. Hogan, L. Posthuma, M. Kavand, Y. Lyu, M. Sherwin, C. Brown, M. J. Graf, L. Balents, S. D. Wilson, arXiv:1901.09408 (2019).
\bibitem{micheal} M. McGuire, Q. Zheng, J. Yan and B. Sales, Phys. Rev. B, 99, 214402 (2019).





\end{thebibliography}
\end{document}